\documentclass[runningheads]{llncs}
\usepackage[utf8]{inputenc} %
\usepackage[T1]{fontenc}    %
\usepackage[per-mode=symbol]{siunitx}
\usepackage{graphicx} %
\usepackage{amsmath} %
\usepackage{amssymb}  %
\usepackage{color}
\usepackage{bm}
\usepackage{nicefrac}
\usepackage{subcaption}
\usepackage{placeins}
\usepackage{pgfplots}
\pgfplotsset{compat=1.9}
\usepgfplotslibrary{polar}

\begin{document}
	\title{Local Communication Protocols for Learning Complex Swarm Behaviors with Deep Reinforcement Learning}%
	\titlerunning{Local Communication Protocols}

	\author{Maximilian H\"uttenrauch\inst{1} \and
	Adrian \v{S}o\v{s}i\'{c}\inst{2} \and
	Gerhard Neumann\inst{1}}
	\authorrunning{H\"uttenrauch et al.}
	\institute{School of Computer Science, University of Lincoln, Lincoln, UK \email{\{mhuettenrauch,gneumann\}@lincoln.ac.uk} \and
	Department of Electrical Engineering, Technische Universit\"at Darmstadt,
	Darmstadt, Germany
	\email{adrian.sosic@spg.tu-darmstadt.de}\\
	}
	\index{H\"uttenrauch, Maximilian}
	\index{\v{S}o\v{s}i\'{c}, Adrian}
	\index{Neumann, Gerhard}
	\maketitle              %
	\begin{abstract}
	Swarm systems constitute a challenging problem for reinforcement learning (RL) as the algorithm needs to learn decentralized control policies that can cope with limited local sensing and communication abilities of the agents.
	While it is often difficult to directly define the behavior of the agents, simple communication protocols can be defined more easily using prior knowledge about the given task.
	In this paper, we propose a number of simple communication protocols that can be exploited by deep reinforcement learning to find decentralized control policies in a multi-robot swarm environment.
	The protocols are based on histograms that encode the local neighborhood relations of the agents and can also transmit task-specific information, such as the shortest distance and direction to a desired target.
	In our framework, we use an adaptation of Trust Region Policy Optimization to learn complex collaborative tasks, such as formation building and building a communication link.
	We evaluate our findings in a simulated 2D-physics environment, and compare the implications of different communication protocols.

	\end{abstract}
	
	\section{Introduction}
	\label{sec:intro}
	\noindent Nature provides many examples where the performance of a collective of limited beings exceeds the capabilities of one individual.
	Ants transport prey of the size no single ant could carry, termites build nests of up to nine meters in height, and bees are able to regulate the temperature of a hive.
	Common to all these phenomena is the fact that each individual has only basic and local sensing of its environment and limited communication capabilities to its neighbors.
	
	Inspired by these biological processes, swarm robotics \cite{BAYINDIR2016292,alonso2016distributed,chen2013strategy} tries to emulate such complex behavior with a collective of rather simple entities.
	Typically, these robots have limited movement and communication capabilities and can sense only a local neighborhood of their environment, such as distances and bearings to neighbored agents.
	Moreover, these agents have limited memory systems, such that the agents can only access a short horizon of their perception.
	As a consequence, the design of control policies that are capable of solving complex cooperative tasks becomes a non-trivial problem.
	
	In this paper, we want to learn swarm behavior using deep reinforcement learning \cite{schulman2015trust,mnih2015human,schulman2017proximal,teh2017distral,GuLilGhaTurLev17} based on the locally sensed information of the agents such that the desired behavior can be defined by a reward function instead of hand-tuning controllers of the agents.
	Swarm systems constitute a challenging problem for reinforcement learning as the algorithm needs to learn decentralized control policies that can cope with limited local sensing and communication abilities of the agents.
	
	Most collective tasks require some form of active cooperation between the agents.
	For efficient cooperation, the agents need to implement basic communication protocols such that they can transmit their local sensory information to neighbored agents.
	Using prior knowledge about the given task, simple communication protocols can be defined much more easily than directly defining the behavior.
	In this paper, we propose and evaluate several communication protocols that can be exploited by deep reinforcement learning to find decentralized control policies in a multi robot swarm environment. 
	
	Our communication protocols are based on local histograms that encode the neighborhood relation of an agent to other agents and can also transmit task-specific information such as the shortest distance and direction to a desired target.
	The histograms can deal with the varying number of neighbors that can be sensed by a single agent depending on its current neighborhood configuration.
	These protocols are used to generate high dimensional observations for the individual agents that is in turn exploited by deep reinforcement learning to efficiently learn complex swarm behavior.
	In particular, we choose an adaptation of Trust Region Policy Optimization \cite{schulman2015trust} to learn decentralized policies.

	In summary, our method addresses the emerging challenges of %
	decentralized swarm control in the following way:
	\begin{enumerate}
		\item \textbf{Homogeneity:} explicit sharing of policy parameters between the agents
		\item \textbf{Partial observability:} efficient processing of action-observation histories through windowing and parameter sharing
		\item \textbf{Communication:} usage of histogram-based communication protocols over simple features
	\end{enumerate}
	To demonstrate our approach, we formulate two cooperative learning tasks in a simulated swarm environment.
	The environment is inspired by the Colias robot \cite{Arvin2014}, a modular platform with two wheel motor-driven movement and various sensing systems.
	
	\paragraph*{Paper Outline}
	In Section~\ref{sec:background}, we review the concepts of Trust Region Policy Optimization and describe our problem domain. 
	In Section~\ref{sec:method}, we show in detail how we tackle the challenges of modeling observations and the policy in the partially observable swarm context, and how to adapt Trust Region Policy Optimization to our setup.
	In Section~\ref{sec:experiments}, we present the model and parameters of our agents and introduce two tasks on which we evaluate our proposed observation models and policies. %

	\section{Background}
	\label{sec:background}
	\noindent In this section, we provide a short summary of Trust Region Policy Optimization and formalize our learning problem domain.
	
	\subsection{Trust Region Policy Optimization}
	\noindent 
	Trust Region Policy Optimization (TRPO) is an algorithm to optimize control policies in single-agent reinforcement learning problems \cite{schulman2015trust}.
	These problems are formulated as Markov decision processes (MDP) which are compactly written as a tuple $\langle \mathcal{S}, \mathcal{A}, P, R, \gamma \rangle$.
	In an MDP, an agent chooses an action $a \in {\mathcal{A}}$ via some policy~$\pi(a \mid s)$, based on its current state $s \in \mathcal{S}$, and progresses to state $s'\in\mathcal{S}$ according to a transition function $P(s' \mid s, a)$.
	After each step, the agent is assigned a reward $r = R(s, a)$, provided by a reward function $R$ which judges the quality of its decision. The goal of the agent is to find a policy which maximizes the expected cumulative reward $\mathbb{E}[\sum_{k=t}^{\infty} \gamma^{k-t} R(s_k, a_k)]$, discounted by factor $\gamma$, achieved over a certain period of time.
	
	In TRPO, the policy is parametrized by a parameter vector~$\theta$ containing weights and biases of a neural network. In the following, we denote this parameterized policy as $\pi_\theta$. %
	The reinforcement learning objective is expressed as finding a new policy that maximizes the expected advantage function of the current policy, i.e., ${J^{\text{TRPO}} = \mathbb{E}\left[\frac{\pi_{\theta}}{\pi_{\theta_{\text{old}}}} \hat{A}(s, a)\right]},$ where $\hat{A}$ is an estimate of the advantage function of the current policy $\pi_{\textrm{old}}$ which is defined as %
	$\hat{A}(s,a) = Q^{\pi_{\textrm{old}}}(s,a)-V^\pi_{\textrm{old}}(s)$. Herein, state-action value function $Q^{\pi_{\textrm{old}}}(s,a)$ is typically estimated by a single trajectory rollout while for the value function $V^{\pi_{\textrm{old}}}(s)$ rather simple baselines are used that are fitted to the monte-carlo returns.
	The objective is to be maximized subject to a fixed constraint on the Kullback-Leibler (KL) divergence of the policy before and after the parameter update,	which ensures the updates to the new policy's parameters $\theta$ are bounded, in order to avoid divergence of the learning process.
	The overall optimization problem is summarized as
	\begin{equation*}
		\begin{aligned}
		& \underset{\theta}{\text{maximize}}
		& & \mathbb{E}\left[\frac{\pi_{\theta}}{\pi_{\theta_{\text{old}}}}\hat{A}(s, a)\right] \\
		& \text{subject to}
		& & \mathbb{E}[D_{\text{KL}}(\pi_{\theta_{\text{old}}}|| \pi_{\theta})] \leq \delta.
		\end{aligned}
	\end{equation*}
	The problem is approximately solved using the conjugate gradient optimizer after linearizing the objective and quadratizing the constraint.
	
	\subsection{Problem Domain}
	\label{sec:prob_state}
	\noindent
	Building upon the theory of single-agent reinforcement learning, we can now formulate the problem domain for our swarm environments.
	Because of their limited sensory input, each agent can only obtain a local observation $o$ from the vicinity of its environment.
	We formulate the swarm system as a swarm MDP (see \cite{Sosic2016} for a similar definition) which can be seen as a special case of a decentralized partially observed Markov decision process (Dec-POMDP) \cite{Oliehoek2013}.
	An agent in the swarm MDP is defined as a tuple $\mathbb{A} = \langle \mathcal{S}, \mathcal{O}, \mathcal{A}, O\rangle$, where, $\mathcal{S}$ is a set of local states, $\mathcal{O}$ is the space of local observations, and ${\mathcal{A}}$ is a set of local actions for each agent.
	The observation model $O(o|s,i)$ defines the observation probabilities for agent $i$ given the global state $s$.
	Note that the system is invariant to the order of the agents, i.e., given the same local state of two agents, the observation probabilities will be the same.
	The swarm MDP is then defined as $\langle N, \mathcal{E}, \mathbb{A}, P, R\rangle$, where $N$ is the number of agents, $\mathcal{E}$ is the global environment state consisting of all local states $\mathcal{S}^N $ of the agents and possibly of additional states of the environment, and $P: \mathcal{S}^N \times \mathcal{S}^N \times \mathcal{A}^N \rightarrow [0, \infty)$ is the transition density function.
	Each agent maintains a truncated history $h_t^i=(a_{t - \eta}^i, o_{t - \eta + 1}^i, \dots, a_{t-1}^i, o_t^i)$ of the current and past observations $o^i \in \mathcal{O}$ and actions $a^i \in \mathcal{A}$ of length $\eta$. 
	All swarm agents are assumed to be identical and therefore use the same distributed policy~$\pi$ (now defined as $\pi(a \mid h)$) which yields a sample for the action of each agent given its current history of actions and observations.
	The reward function $R$ of the swarm MDP depends on the {\em global state} and, optionally, all actions of the swarm agents, i.e., $R: \mathcal{S}^N \times \mathcal{A}^N \rightarrow \mathbb{R}$. 
	Instead of considering only one single agent, we consider multiple agents of the same type, which interact in the same environment.
	The global system state is in this case comprised of the local states of all agents and additional attributes of the environment.
	The global task of the agents is encoded in the reward function $R(\bm{s},\bm{a})$, where we from now on write $\bm{a}$ to denote the joint action vector of the whole swarm.
	
	\subsection{Related Work}
	\noindent 
	A common approach to program swarm robotic systems is by extracting rules from the observed behavior of their natural counterparts.
	Kube et al \cite{Kube1998}, for example, investigate the cooperative prey retrieval of ants to infer rules on how a swarm of robots can fulfill the task of cooperative box-pushing.
	Similar work can be found e.g.\ in \cite{Martinoli2004}, \cite{Hoff2010}, \cite{Nouyan2009}.
	However, %
	extracting these rules can be tedious and the complexity of the tasks that we can solve %
	via explicit programming is limited.
	More examples of rule based behavior are found in \cite{chen2013strategy} where a group of swarming robots transports an object to a goal.
	Further comparable work can be found in \cite{Correll2011} for aggregation, \cite{Moeslinger2010} for flocking, or \cite{Goldberg2000} for foraging.
	
	In deep RL, currently, there are only few approaches tackling the multi-agent problem.
	One of these approaches can be found in \cite{LoweWTHAM17}, where the authors use a variation of the deep deterministic policy gradient algorithm \cite{Lillicrap2015} to learn a centralized Q-function for each policy, which, as a downside, leads to a linear increase in dimensionality of the joint observation and action spaces therefore scales poorly.
	Another algorithm, tackling the credit assignment problem, can be found in \cite{Foerster17}.
	Here, a baseline of other agents' behavior is subtracted from a centralized critic to reason about the quality of a single agent's behavior.
	However, this approach is only possible in scenarios with discrete action spaces since it requires marginalization over the agents' action space.
	Finally, a different line of work concerning the learning of communication models between agents can be found in \cite{FoersterAFW16a}.	
	
	\section{Multi-Agent Learning with Local Communication Protocols}
	\label{sec:method}
	\noindent In this section, we introduce different communication protocols based on neighborhood histograms that can be used in combination to solve complex swarm behaviors.
	Our algorithm relies on deep neural network policies of special architecture that can exploit the structure of the high-dimensional observation histories.
	We present this network model and subsequently discuss small adaptations we had to make to the TRPO algorithm in order to apply it to this cooperative multi-agent setting.    
	
	\begin{figure}[t]
	\centering
	
	\resizebox{10cm}{3cm}{
		
		\begin{tikzpicture}[inner sep=0mm]
		
		\definecolor{agentGreen}{RGB}{55, 124, 33}
		\definecolor{sensor}{RGB}{205, 222, 199}
		\definecolor{sensorBoundary}{RGB}{168, 197, 157}
		\newcommand{\fac}{0.25}
		\tikzstyle{agent}=[circle, draw=agentGreen, fill=agentGreen, inner sep=0, minimum size=2mm]
		
		\node [anchor=west] at (2,0) {
			\subcaptionbox{\Large local agent configuration \label{fig:agent_config}}{
				\begin{tikzpicture}[anchor=center]
				\filldraw [fill=sensor, draw=sensorBoundary] (0,0)  -- ++ (67.5:8cm*\fac) arc[green, radius = 8cm*\fac, start angle= 67.5, end angle=112.5] -- (0,0);
				\foreach \angle in {22.5,67.5,...,360}	
				\draw (0,0) -- (\angle:8cm*\fac);
				\foreach \radius in {0, 2, ...,8}	
				\draw (0,0) circle (\radius cm*\fac);
				
				\node [agent, minimum size=3mm] at (0,0) {};
				\node [agent] at (45:1.5 cm*\fac) {};
				\node [agent] at (10:5.5 cm*\fac) {};
				\node [agent] at (-95:3.2cm*\fac) {};
				\node [agent] at (120:6.6cm*\fac) {};
				\node [agent] at (140:3.5cm*\fac) {};
				\node [agent] at (-140:6.3cm*\fac) {};
				\node [agent] at (-45:7.4cm*\fac) {};
				\node [agent] at (75:5.4cm*\fac) {};
				\end{tikzpicture}
			}};
			
			\node at (\textwidth/2 + 3cm,1) {
				\subcaptionbox{\Large range histogram \label{fig:dist_hist}}{
					\begin{tikzpicture}[scale=0.5]
					\foreach \x in {0,2,...,8}
					\draw (\x,0) -- (\x,1cm);
					\foreach \y in {0,...,1}
					\draw (0,\y) -- (8cm,\y);
					
					\node [agent] at (1,0.5) {};
					\node [agent] at (4.5,0.7) {};
					\node [agent] at (2.5,0.7) {};
					\node [agent] at (7.5,0.3) {};
					\node [agent] at (3.5,0.3) {};
					\node [agent] at (6.5,0.7) {};
					\node [agent] at (7,0.5) {};
					\node [agent] at (5.5,0.3) {};
					\node at (4, -0.75) {\Large $d$};
					\end{tikzpicture}		
				}};
				
				\node at (\textwidth/2 + 3cm,-1.25) {
					\subcaptionbox{\Large bearing histogram \label{fig:bear_hist}}{
						\begin{tikzpicture}[scale=0.5]
						\draw [fill=sensor] (3,0) rectangle (4,1);
						
						\foreach \x in {0,...,8}
						\draw (\x,0) -- (\x,1cm);
						\foreach \y in {0,...,1}
						\draw (0,\y) -- (8cm,\y);
						
						\node [agent] at (4.5,0.5) {};
						\node [agent] at (5.5,0.5) {};
						\node [agent] at (7.5,0.5) {};
						\node [agent] at (2.7,0.3) {};
						\node [agent] at (2.3,0.7) {};
						\node [agent] at (0.5,0.5) {};
						\node [agent] at (6.5,0.5) {};
						\node [agent] at (3.5,0.5) {};
						\node at (4, -0.75) {\Large $\phi$};
						\end{tikzpicture}
					}};
					
					\node [anchor=west] at (\textwidth ,0) {
						\subcaptionbox{\Large joint histogram \label{fig:joint_hist}}{
							\begin{tikzpicture}[scale=0.75, anchor=center]
							\draw [fill=sensor] (3,0) rectangle (4,4);
							
							\foreach \x in {0,...,8}
							\draw (\x,0) -- (\x,4cm);
							\foreach \y in {0,...,4}
							\draw (0,\y) -- (8cm,\y);
							
							\node [agent] at (4.5,0.5) {};
							\node [agent] at (3.5,2.5) {};
							\node [agent] at (5.5,2.5) {};
							\node [agent] at (7.5,1.5) {};
							\node [agent] at (2.5,1.5) {};
							\node [agent] at (2.5,3.5) {};
							\node [agent] at (0.5,3.5) {};
							\node [agent] at (6.5,3.5) {};
							\node at (-0.5, 2) {\Large $d$};
							\node at (4, -0.75) {\Large $\phi$};
							\end{tikzpicture}
						}};
						
			\end{tikzpicture}
		}
		\caption{This Figure shows an illustration of the histogram-based observation model. Figure \ref{fig:agent_config} shows an agent in the center of a circle whose neighborhood relations are to be captured by the histogram representation. The shaded green area is highlighted as a reference for Figures \ref{fig:bear_hist} and \ref{fig:joint_hist}. Figure \ref{fig:dist_hist} hereby shows the one dimensional histogram of agents over the neighborhood range $d$ into four bins, whereas Figure \ref{fig:bear_hist} shows the histogram over the bearing angles $\phi$ into eight bins. Figure \ref{fig:joint_hist} finally shows the two dimensional joint histogram over range and bearing.}
		\label{fig:histogramIllustration}
	\end{figure}
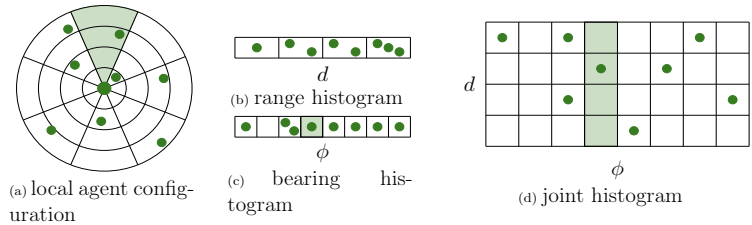
	\subsection{Communication Protocols}
	\noindent Our communication protocols are based on histograms that can either encode neighborhood relations or distance relations to different points of interest. 
	\subsection*{Neighborhood Histograms}
	\noindent The individual agents can observe distance and bearing to neighbored agents if they communicate with this agent.
	We assume that the agents are constantly sending a signal, such that neighbored agents can localize the sources.
	The arising neighborhood configuration is an important source of information and can be used as observations of the individual agents. 
	One of the arising difficulties in this case is to handle changing number of neighbors which would result in a variable length of the observation vector.
	Most policy representations, such as neural networks, expect a fixed input dimension.
	
	One possible solution to this problem is to allocate a fixed number of neighbor relations for each agent.
	If an agent experiences fewer neighborhood relations, standard values could be used such as a very high distance and 0 bearing. 
	However, such an approach comes with several drawbacks.
	First of all, the size of the resulting representation scales linearly with the number of agents in the system and so does the number of parameters to be learned.
	Second, the execution of the learned policy will be limited to scenarios with the exact same number of agents as present during training.
	Third, a fixed allocation of the neighbor relation inevitably destroys the homogeneity of the swarm, since the agents are no longer treated interchangeably.
	In particular, using a fixed allocation rule requires that the agents must be able to discriminate between their neighbors, which might not even be possible in the first place.
	
	To solve these problems, we propose to use \textit{histograms over observed neighborhood relations}, e.g., distances and bearing angles.
	Such a representation inherently respects the agent homogeneity and naturally comes with a fixed dimensionality.
	Hence, it is the canonical choice for the swarm setting. 
	For our experiments, we consider two different types of representations: 1) concatenated one-dimensional histograms of distance and bearing and 2)~multidimensional histograms. Both types are illustrated in Figure~\ref{fig:histogramIllustration}.
	The one-dimensional representation has the advantage of scalability, as it grows linearly with the number of features.
	The downside is that potential dependencies between the features are completely ignored.
	
	\subsection*{Shortest Path Partitions}
	In many applications, it is important to transmit the location of a point of interest to neighbored agents that can currently not observe this point due to their limited sensing ability. 
	
	We assume that an agent can observe bearing and distance to a point of interest if it is within its communication radius.
	The agent then transmits the observed distance to other agents.
	Agents that can not see the point of interest might in this case observe a message from another agent containing the distance to the point of interest.
	The distance of the sending agent is added to the received distance to obtain the distance to the point of interest if we would use the sending agent as a via point.
	Each agent might now compute several of such distances and transmits the minimum distance it has computed to indicate the length of the shortest path it has seen.   
	
	The location of neighbored agents including their distance of the shortest path information is important knowledge for the policy, e.g.\ for navigating to the point of interest.
	Hence, we adapt the histogram representation.
	Each partition now contains the minimum received shortest path distance of an agent that is located in this position. 
	
	\subsection{Weight Sharing for Policy Networks}
\begin{figure}[h]
		\begin{minipage}[c]{0.6\textwidth}
			\includegraphics[width=0.8\columnwidth]{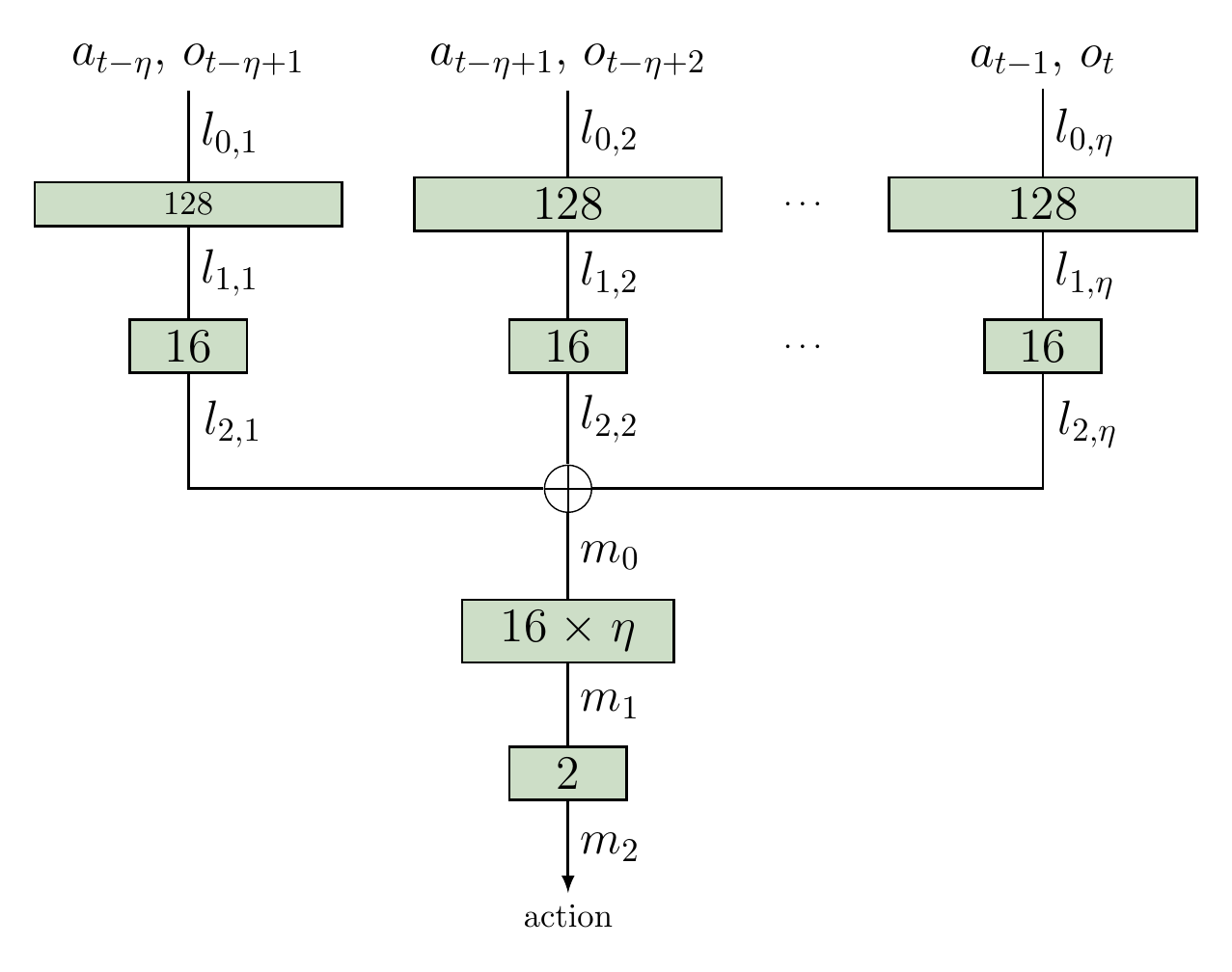}
		\end{minipage}\hfill
		\begin{minipage}[c]{0.4\textwidth}
			\caption{This diagram shows a model of our proposed policy with three hidden layers. The numbers inside the boxes denote the dimensionalities of the hidden layers. The plus sign denotes concatenation of vectors.}
			\label{fig:policy_model}
		\end{minipage}
	\end{figure}

	\noindent The policy maps sequences of past actions and observations to a new action.
	We use histories of a fixed length as input to our policy and a feed-forward deep neural network as architecture.
	To cope with such high input dimensionality, we propose a weight sharing approach.
	Each action-observation pair in an agent's history is first processed independently with a network using the same weights.
	After this initial reduction in dimensionality, the hidden states are concatenated in a subsequent layer and finally mapped to an output.
	The homogeneity of agents is achieved by using the same set of parameters for all policies.
	A diagram of the architecture is shown in Figure~\ref{fig:policy_model}.

	\subsection{Adaptations to TRPO}
	\noindent In order to apply TRPO to our multi-agent setup, some small changes to the original algorithm have to be made, similar to the formulation of \cite{Gupta2017}.
	First, since we assume homogeneous agents, we can have one set of parameters of the policy shared by all agents.
	Since the agents cannot rely on the global state, the advantage function is redefined as $A(h, a)$.
	In order to estimate this function, each agent is assigned the same global reward $r$ in each time step and all transitions are treated as if they were executed by a single agent.
	
	\section{Experimental Setup}
	\label{sec:experiments}	
	\noindent In this section, we briefly discuss the used model and state representation of a single agent.
	Subsequently, we describe our two experimental setups and the policy architecture used for the experiments. 
	
	\subsection{Agent Model}
	\noindent The local state of a single agent is modeled by its 2D position and orientation, i.e., $s^i = [x^i, y^i, \phi^i] \in \mathcal{S} = \{[x, y, \phi] \in \mathbb{R}^3 : 0 \leq x \leq x_{\text{max}},\ 0 \leq y \leq y_{\text{max}}, \ 0 \leq \phi \leq 2\pi \}$.
	The robot can only control the speed of its wheels.
	Therefore, we apply a force to the left and right side of the agent, similarly to %
	the wheels of the real robot. 
	Our model of a single agent is inspired by the Colias robot (a detailed description of the robot specifications can be found in \cite{Arvin2014}), but the underlying principles can be straightforwardly applied to other swarm settings with limited observations.
	Generally, our observation model is comprised of the sensor readings of the short and long range IR sensors (later denoted as 'sensor' in the evaluations).
	Furthermore, we augment this observation representation with the communication protocols developed in the following section.
	Our simulation is using a 2D physics engine (Box2D), allowing for correct physical interaction of the bodies of the agents.
	
	\subsection{Tasks}
	\noindent The focus of our experiments is on tasks where agents need to collaborate to achieve a common goal.
	For this purpose, we designed the following two scenarios:
	
	\subsection*{Task 1: Building a Graph}
	\noindent In the first task, the goal of the agents is to find and maintain a certain distance to each other. 
	This kind of behavior is required, for example, in surveillance tasks, where a group of autonomous agents needs to maximize the coverage of a target area while maintaining their communication links.
	We formulate the task as a graph problem, where the agents (i.e.\ the nodes) try to maximize the number of active edges in the graph.
	Herein, an edge is considered active whenever the distance between the corresponding agent lies in certain range.
	The setting is visualized in Figure~\ref{fig:edge_example}.
	In our experiment, we provide a positive reward for each edge in a range between $\SI{10}{\cm}$ and $\SI{16}{\cm}$, and further give negative feedback for distances smaller than $\SI{7}{\cm}$.
	Accordingly, the reward function is
	\begin{align}
	R(\bm{s}, \bm{a}) = \sum_{i=1}^{M} \sum_{m>i}^{M} \bm{1}_{[\SI{0.1}{\meter}, ~ \SI{0.16}{\meter}]} (d_m^i) - 5 \sum_{i=1}^{M} \sum_{m>i}^{M} \bm{1}_{[\SI{0}{\meter}, ~ \SI{0.07}{\meter}]} (d_m^i),
	\end{align}
	where $d_m^i = \sqrt{(x_i -x_m)^2 + (y_i - y_m)^2}$ denotes the Euclidean distance between the centers of agent $i$ and agent $m$ and
	\begin{align*}
	\bm{1}_{[a,b]}(x) &= \begin{cases}
	1 & \text{if \,} x \in [a, b],\\
	0 & \text{else}
	\end{cases}
	\end{align*}
	is an indicator function. %
	Note that we omit the dependence of $d_m^i$ on the system state $s$ to keep the notion simple.
	
	\subsection*{Task 2: Establishing a Communication Link}
\begin{figure}[h]
		\centering
		\begin{subfigure}[t]{0.49\textwidth}
			\centering
			\includegraphics[width=0.7\columnwidth]{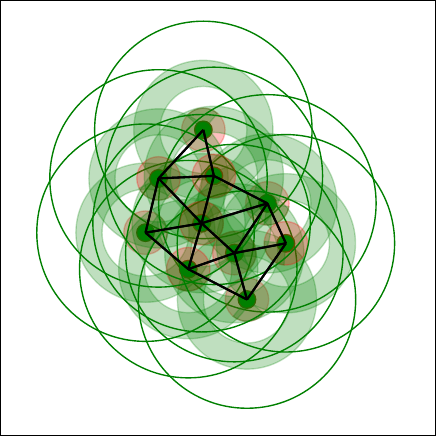}
			\caption{edge task}
			\label{fig:edge_example}
		\end{subfigure}
		\hfill
		\begin{subfigure}[t]{0.49\textwidth}
			\centering
			\includegraphics[width=0.7\columnwidth]{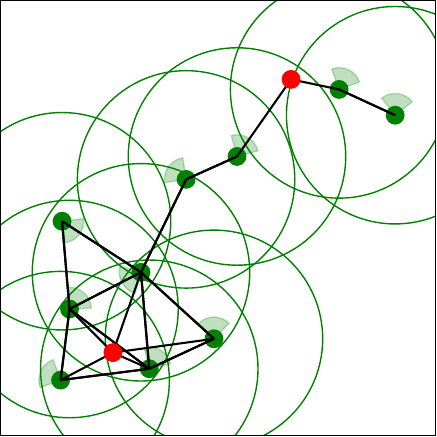}
			\caption{link task}
			\label{fig:link_example}
		\end{subfigure}
		\caption{Illustration of the two cooperative tasks used in this paper. The green dots represent the agents, where the green ring segments located next to the agents indicate the short range IR front sensors. The outer green circles illustrate the maximum range in which distances\,/\,bearings to other agents can be observed, depending on the used observation model. \textbf{(a)~Edge task:} The red rings show the penalty zones where the agents are punished, the outer green rings indicate the zones where legal edges are formed. \textbf{(b)~Link task:} The red dots correspond to the two points that need to be connected by the agents.
		}
	\end{figure}

	\noindent The second task adds another layer of difficulty.
	While maintaining a network, the agents have to locate and connect two randomly placed points in the state space.
	A link is only established successfully if there are communicating agents connecting the two points.
	Figure~\ref{fig:link_example} shows an example with an active link spanned by three agents between the two points.
	The task resembles the problem of establishing a connection between two nodes in a wireless ad hoc network \cite{basu2004movement,witkowski2008ad}.
	In our experiments, the distance of the two points is chosen to be larger than $\SI{75}{\cm}$, requiring at least three agents to bridge the gap in between.
	The reward is determined by the length of the shortest distance between the two points $d_{\text{opt}}$ (i.e.\ a straight line) and the length of the shortest active link $d_{\text{sp}}$ spanned by the agents,
	\begin{equation*}
	R(\bm{s}, \bm{a}) = \begin{cases}
	\frac{d_{\text{opt}}}{d_{\text{sp}}} & \text{if link is established} \\
	0 & \text{otherwise.}
	\end{cases}
	\end{equation*}
	In this task, we use the shortest path partitions as communication protocol.
	Each agent communicates the shortest path it knows to both points of interests, resulting in two 2-D partitions that are used as observation input for a single time step.

	\subsection{Policy Architecture}
	\noindent We decided for a policy model with three hidden layers.
	The first two layers process the observation-action pairs $(a_{k-1}, o_{k})$ of each timestep in a history individually and map it into hidden layers of size 128 and 16.
	The output of the second layer is then concatenated to form the input of the third hidden layer which eventually maps to the two actions for the left and right motor.
	
	\section{Results}
	\noindent We evaluate each task in a standardized environment of size $\SI{1}{\m}\times\SI{1}{\m}$ where we initialize ten agents randomly in the scene.
	Of special interest is how the amount of information provided to the agents affects the overall system performance. 
	Herein, we have to keep in mind the general information-complexity trade-off, i.e., high-dimensional local observations generally provide more information about the global system state but, at the same time, result in a more complex learning task. Recall that the information content is mostly influenced by two factors: 1) the length of the history, and 2) the composition of the observation.
	
	\subsection{Edge Task}
	\noindent %
	First, we evaluate how the history length $\eta$ affects the system performance.
	Figure~\ref{fig:edge_history_ws} shows an evaluation for $\eta = \{2, 4, 8\}$ and a weight sharing policy using a two-dimensional histogram over distances and bearings.
	Interestingly, we observe that longer observation histories do not show an increase in the performance.
	Either the increase in information could not counter the effect of increased learning complexity, or a history length of $\eta = 2$ is already sufficient to solve the task.
	We use these findings and set the history length to $\eta = 2$ for the remainder of the experiments.
	
	Next, we analyze the impact of the observation model. Figure~\ref{fig:edge_obs_models_ws} shows the results of the learning process for different observation modalities.
	The first observation is that, irrespective of the used mode, the agents are able to establish a certain number of edges.
	Naturally, a complete information of distances and bearing yields the best performance.
	However, the independent histogram representation yields comparable results to the two dimensional histogram.
	Again, this is due to the aforementioned complexity trade-off where a higher amount of information makes the learning process more difficult.
	
	\subsection{Link Task}
\begin{figure}[h]
		\begin{subfigure}[t]{0.3\textwidth}
			\centering
			\includegraphics[width=\columnwidth]{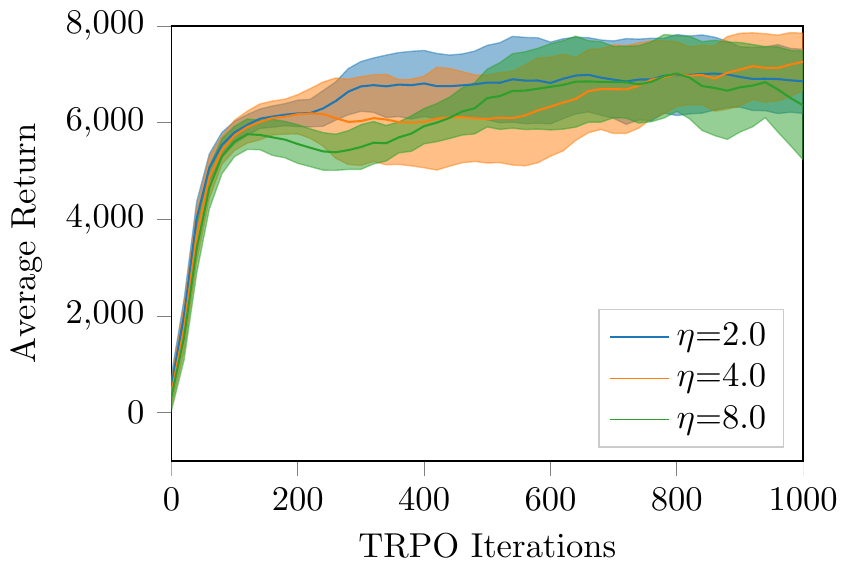}
			\caption{Comparison of different history lengths $\eta$. (2D histogram)}
			\label{fig:edge_history_ws}
		\end{subfigure}
		\hfill
		\begin{subfigure}[t]{0.3\textwidth}
			\centering
			\includegraphics[width=\columnwidth]{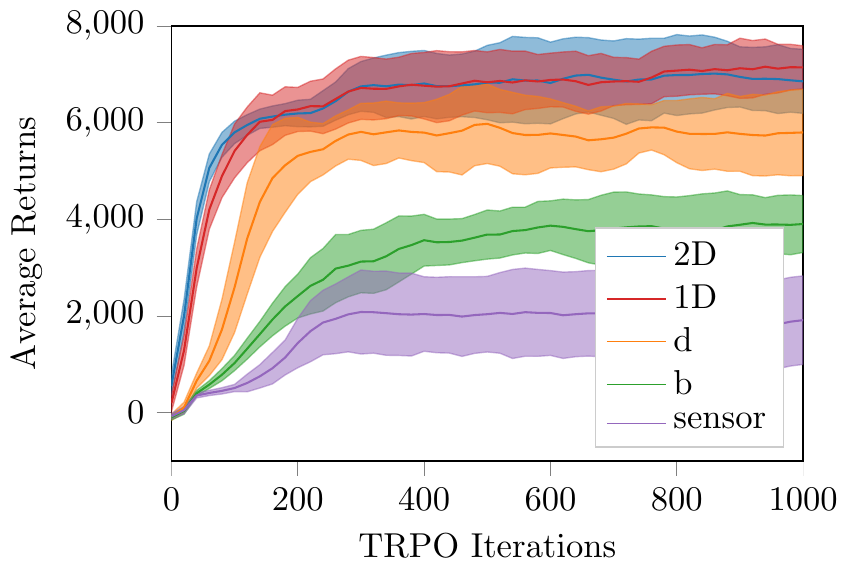}
			\caption{Edge task: Comparison of different observation models. ($\eta=2$)}
			\label{fig:edge_obs_models_ws}
		\end{subfigure}
		\hfill
		\begin{subfigure}[t]{0.3\textwidth}
			\centering
			\includegraphics[width=\columnwidth]{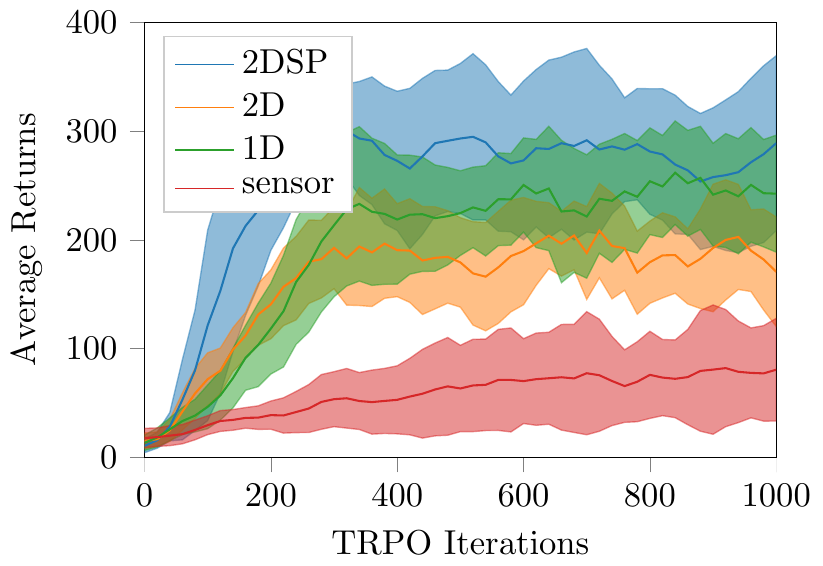}
			\caption{Link task: Comparison of different observation models. ($\eta=2$)}
			\label{fig:chain_obs_models_ws}
		\end{subfigure}
		\caption{Learning curves for (a), (b)~the edge task and (c)~the link task. The curves show the mean values of the average undiscounted return of an episode (i.e. the sum of rewards of one episode, averaged over the number of episodes for one learning iteration) over the learning process plus\,/\,minus one standard deviation, computed from eight learning trials. Intuitively, the return in the edge task corresponds to the number of edges formed during an episode of length 500 steps. In the link task, it is a measure for the quality of the link. 
		\textbf{Legend:} 2DSP: two dimensional histogram over shortest paths, 2D: two-dimensional histogram over distances and bearings, 1D: two independent histograms over distances and bearing, d: distance only histogram, b: bearing only histogram, sensor: no histogram.
		}
	\end{figure}

	\noindent We evaluate the link task with raw sensor measurements, count based histograms over distance and bearing, and the more advanced shortest path histograms over distance and bearing. 
	Based on the findings of the edge task we keep the history length at $\eta = 2$.
	Figure~\ref{fig:chain_obs_models_ws} shows the results of the learning process where each observation model was again tested and averaged over 8 trials.
	Since at least three agents are necessary to establish a link between the two points, the models without shortest path information struggle to reliably establish the connection.
	Their only chance is to spread as wide as possible and, thus, cover the area between both points.
	Again, it is interesting to see that independent histograms over counts seem to be favorable over the 2D histogram.
	However, both versions are surpassed by the 2D histogram over shortest paths which yields information about the current state of the whole network of agents, currently connected to each of the points. 

	\section{Conclusions and Future Work}
	\noindent In this paper, we demonstrated that histograms over simple local features can be an effective way for processing information in robot swarms.
	The central aspect of this new model is its ability to handle arbitrary system sizes without discriminating between agents, which makes it perfectly suitable to the swarm setting where all agents are identical and the number of agents in the neighborhood varies with time.
	We use these protocols and an adaptation of TRPO for the swarm setup to learn cooperative decentralized control policies for a number of challenging cooperative task.
	The evaluation of our approach showed that this histogram-based model leads the agents to reliably fulfill the tasks.
	
	Interesting future directions include, for example, the learning of an explicit communication protocol.
	Furthermore, we expect that assigning credit to agents taking useful actions should speedup our learning algorithm.
	
	\bigskip
	\subsubsection*{Acknowledgments. }
	The research leading to these results has received funding from EPSRC under grant agreement EP/R02572X/1 (National Center for Nuclear Robotics).
	Calculations for this research were conducted on the Lichtenberg high performance computer of the TU Darmstadt.

	\clearpage
	\bibliographystyle{splncs04}
	\bibliography{ants}

\begin{thebibliography}{10}
\providecommand{\url}[1]{\texttt{#1}}
\providecommand{\urlprefix}{URL }
\providecommand{\doi}[1]{https://doi.org/#1}

\bibitem{alonso2016distributed}
Alonso-Mora, J., Montijano, E., Schwager, M., Rus, D.: Distributed multi-robot
  formation control among obstacles: A geometric and optimization approach with
  consensus. In: Proceedings of the IEEE International Conference on Robotics
  and Automation. pp. 5356--5363 (2016)

\bibitem{Arvin2014}
Arvin, F., Murray, J., Zhang, C., Yue, S.: Colias: An autonomous micro robot
  for swarm robotic applications. International Journal of Advanced Robotic
  Systems  \textbf{11}(7), ~113 (2014)

\bibitem{basu2004movement}
Basu, P., Redi, J.: Movement control algorithms for realization of
  fault-tolerant ad hoc robot networks. IEEE Network  \textbf{18}(4),  36--44
  (2004)

\bibitem{BAYINDIR2016292}
Bayındır, L.: A review of swarm robotics tasks. Neurocomputing
  \textbf{172}(C),  292 -- 321 (2016)

\bibitem{chen2013strategy}
Chen, J., Gauci, M., Gro{\ss}, R.: A strategy for transporting tall objects
  with a swarm of miniature mobile robots. In: Proceedings of the IEEE
  International Conference on Robotics and Automation. pp. 863--869 (2013)

\bibitem{Correll2011}
Correll, N., Martinoli, A.: Modeling and designing self-organized aggregation
  in a swarm of miniature robots. The International Journal of Robotics
  Research  \textbf{30}(5),  615--626 (2011)

\bibitem{FoersterAFW16a}
Foerster, J., Assael, Y.M., de~Freitas, N., Whiteson, S.: Learning to
  communicate with deep multi-agent reinforcement learning. In: Advances in
  Neural Information Processing Systems 29, pp. 2137--2145 (2016)

\bibitem{Foerster17}
Foerster, J., Farquhar, G., Afouras, T., Nardelli, N., Whiteson, S.:
  Counterfactual multi-agent policy gradients. arXiv:1705.08926  (2017)

\bibitem{Goldberg2000}
Goldberg, D., Mataric, M.J.: Robust behavior-based control for distributed
  multi-robot collection tasks  (2000)

\bibitem{GuLilGhaTurLev17}
Gu, S., Lillicrap, T., Ghahramani, Z., Turner, R.E., Levine, S.: Q-prop:
  Sample-efficient policy gradient with an off-policy critic. In: Proceedings
  of the 5th International Conference on Learning Representations (2017)

\bibitem{Gupta2017}
Gupta, J.K., Egorov, M., Kochenderfer, M.: Cooperative multi-agent control
  using deep reinforcement learning. In: Proceedings of the Adaptive and
  Learning Agents Workshop (2017)

\bibitem{Hoff2010}
Hoff, N.R., Sagoff, A., Wood, R.J., Nagpal, R.: Two foraging algorithms for
  robot swarms using only local communication. In: Proceedings of the IEEE
  International Conference on Robotics and Biomimetics. pp. 123--130 (2010)

\bibitem{Kube1998}
Kube, C., Bonabeau, E.: Cooperative transport by ants and robots. Robotics and
  Autonomous Systems  \textbf{30}(1),  85 -- 101 (2000)

\bibitem{Lillicrap2015}
Lillicrap, T.P., Hunt, J.J., Pritzel, A., Heess, N., Erez, T., Tassa, Y.,
  Silver, D., Wierstra, D.: Continuous control with deep reinforcement
  learning. arXiv:1509.02971  (2015)

\bibitem{LoweWTHAM17}
Lowe, R., Wu, Y., Tamar, A., Harb, J., Abbeel, P., Mordatch, I.: Multi-agent
  actor-critic for mixed cooperative-competitive environments. arXiv:1706.02275
   (2017)

\bibitem{Martinoli2004}
Martinoli, A., Easton, K., Agassounon, W.: Modeling swarm robotic systems: a
  case study in collaborative distributed manipulation. The International
  Journal of Robotics Research  \textbf{23}(4-5),  415--436 (2004)

\bibitem{mnih2015human}
Mnih, V., Kavukcuoglu, K., Silver, D., Rusu, A.A., Veness, J., Bellemare, M.G.,
  Graves, A., Riedmiller, M., Fidjeland, A.K., Ostrovski, G., et~al.:
  Human-level control through deep reinforcement learning. Nature
  \textbf{518}(7540),  529--533 (2015)

\bibitem{Moeslinger2010}
Moeslinger, C., Schmickl, T., Crailsheim, K.: Emergent flocking with low-end
  swarm robots, pp. 424--431 (2010)

\bibitem{Nouyan2009}
Nouyan, S., Gross, R., Bonani, M., Mondada, F., Dorigo, M.: Teamwork in
  self-organized robot colonies. IEEE Transactions on Evolutionary Computation
  \textbf{13}(4),  695--711 (2009)

\bibitem{Oliehoek2013}
Oliehoek, F.A.: Decentralized POMDPs, pp. 471--503. Springer Berlin Heidelberg
  (2012)

\bibitem{schulman2015trust}
Schulman, J., Levine, S., Moritz, P., Jordan, M., Abbeel, P.: Trust region
  policy optimization. In: Proceedings of the 32nd International Conference on
  Machine Learning. pp. 1889--1897 (2015)

\bibitem{schulman2017proximal}
Schulman, J., Wolski, F., Dhariwal, P., Radford, A., Klimov, O.: Proximal
  policy optimization algorithms. arXiv:1707.06347  (2017)

\bibitem{teh2017distral}
Teh, Y.W., Bapst, V., Czarnecki, W.M., Quan, J., Kirkpatrick, J., Hadsell, R.,
  Heess, N., Pascanu, R.: Distral: robust multitask reinforcement learning.
  arXiv:1707.04175  (2017)

\bibitem{Sosic2016}
\v{S}o\v{S}i\'{c}, A., KhudaBukhsh, W.R., Zoubir, A.M., Koeppl, H.: Inverse
  reinforcement learning in swarm systems. In: Proceedings of the 16th
  Conference on Autonomous Agents and MultiAgent Systems. pp. 1413--1421 (2017)

\bibitem{witkowski2008ad}
Witkowski, U., El~Habbal, M.A.M., Herbrechtsmeier, S., Tanoto, A., Penders, J.,
  Alboul, L., Gazi, V.: Ad-hoc network communication infrastructure for
  multi-robot systems in disaster scenarios. In: Proceedings of the IARP/EURON
  Workshop on Robotics for Risky Interventions and Environmental Surveillance
  (2008)

\end{thebibliography}

\end{document}